\documentclass[aip,jcp,reprint,noshowkeys,superscriptaddress]{revtex4-1}
\usepackage{graphicx,dcolumn,bm,xcolor,microtype,multirow,amscd,amsmath,amssymb,amsfonts,physics,longtable,wrapfig,txfonts,siunitx,float}
\usepackage[version=4]{mhchem}

\usepackage[utf8]{inputenc}
\usepackage[T1]{fontenc}
\usepackage{txfonts}

\usepackage[
	colorlinks=true,
    citecolor=blue,
    breaklinks=true
	]{hyperref}
\newcommand{\ie}{\textit{i.e.}}
\newcommand{\eg}{\textit{e.g.}}
\newcommand{\alert}[1]{\textcolor{black}{#1}}
\usepackage[normalem]{ulem}

\newcommand{\SupMat}{\textcolor{blue}{supplementary material}}
\newcommand{\QP}{\textsc{quantum package}}
\newcommand{\T}[1]{#1^{\intercal}}

\newcommand{\br}{\mathbf{r}}
\newcommand{\bx}{\mathbf{x}}

\newcommand{\Norb}{K}
\newcommand{\Nocc}{O}
\newcommand{\Nvir}{V}

\newcommand{\HF}{\text{HF}}

\newcommand{\ppRPA}{\text{pp-RPA}}
\newcommand{\BSE}{\text{BSE}}
\newcommand{\dBSE}{\text{dBSE}}

\newcommand{\GT}{GT}

\newcommand{\Ec}{E_\text{c}}

\newcommand{\e}[2]{\eps_{#1}^{#2}}

\newcommand{\Om}[2]{\Omega_{#1}^{#2}}

\newcommand{\SO}[1]{\psi_{#1}}

\newcommand{\bO}{\mathbf{0}}

\newcommand{\bA}[2]{\mathbf{A}_{#1}^{#2}}
\newcommand{\bB}[2]{\mathbf{B}_{#1}^{#2}}
\newcommand{\bC}[2]{\mathbf{C}_{#1}^{#2}}
\newcommand{\bX}[2]{\mathbf{X}_{#1}^{#2}}
\newcommand{\bY}[2]{\mathbf{Y}_{#1}^{#2}}

\newcommand{\eps}{\varepsilon}

\newcommand{\cT}{\mathcal{T}}
\DeclareSIUnit\angstrom{\text{\AA}}

\newcommand{\LCPQ}{Laboratoire de Chimie et Physique Quantiques (UMR 5626), Universit\'e de Toulouse, CNRS, UPS, France}
\newcommand{\LPT}{Laboratoire de Physique Th\'eorique, Universit\'e de Toulouse, CNRS, UPS, France}
\newcommand{\ETSF}{European Theoretical Spectroscopy Facility (ETSF)}
\begin{document}	

\title{Static and Dynamic Bethe-Salpeter Equations in the $T$-Matrix Approximation}
\author{Pierre-Fran\c{c}ois \surname{Loos}}
	\email{loos@irsamc.ups-tlse.fr}
	\affiliation{\LCPQ}
\author{Pina \surname{Romaniello}}
	\email{romaniello@irsamc.ups-tlse.fr}
	\affiliation{\LPT}
	\affiliation{\ETSF}

\begin{abstract}
While the well-established $GW$ approximation corresponds to a resummation of the direct ring diagrams and is particularly well suited for weakly-correlated systems, the $T$-matrix approximation does sum ladder diagrams up to infinity and is supposedly more appropriate in the presence of strong correlation.
Here, we derive and implement, for the first time, the static and dynamic Bethe-Salpeter equations when one considers $T$-matrix quasiparticle energies as well as a $T$-matrix-based kernel.
The performance of the static scheme and its perturbative dynamical correction are assessed by computing the neutral excited states of molecular systems.
Comparison with more conventional schemes as well as other wave function methods are also reported.
\alert{Our results suggest that the $T$-matrix-based formalism performs best in few-electron systems where the electron density remains low.}
\end{abstract}

\maketitle

\section{Introduction}
\label{sec:intro}
The $GW$ approximation \cite{Hedin_1965} of many-body perturbation theory \cite{Martin_2016} is becoming a method of choice to target charged excitations (\ie, ionization potentials and electron affinities) in molecular systems. \cite{Aryasetiawan_1998,Onida_2002,Reining_2017,Golze_2019,Bruneval_2021} 
These so-called quasiparticle energies can be experimentally measured from direct and inverse photoemission spectroscopies.
From a more theoretical point of view, $GW$ corresponds to an elegant resummation of all direct ring diagrams from the particle-hole (ph) channel which is particularly justified in the high-density or weakly-correlated regime. \cite{Gell-Mann_1957,Nozieres_1958}
Within the $GW$ approximation, the self-energy --- one of the key quantities of Hedin's equations \cite{Hedin_1965} --- reads
\begin{equation}
\label{eq:SigGW}
	\Sigma^{GW}(1,2) = i G(1,2) W(1,2)
\end{equation}
where $G$ is the one-body Green's function, $W$ is the dynamically-screened Coulomb potential, and, \eg, $1 \equiv (\sigma_1,\br_1,t_1)$ is a composite coordinate gathering spin, space, and time variables.

Alternatives to $GW$ do exist. For example, the $T$-matrix (or Bethe-Goldstone) approximation, first introduced in nuclear physics, \cite{Bethe_1957,Baym_1961,Baym_1962,Danielewicz_1984a,Danielewicz_1984b} then in condensed matter physics, \cite{Liebsch_1981,Bickers_1989,Bickers_1991,Katsnelson_1999,Katsnelson_2002,Zhukov_2005,vonFriesen_2010,Romaniello_2012,Gukelberger_2015,Muller_2019,Friedrich_2019,Biswas_2021} and more recently in quantum chemistry, \cite{Zhang_2017,Li_2021b} sums to infinity the ladder diagrams from the particle-particle (pp) channel and is justified in the low-density or strongly-correlated regime. \cite{Danielewicz_1984a,Danielewicz_1984b,Liebsch_1981,Shepherd_2014}
While the two-point screened interaction $W$ is the cornerstone of $GW$, the $T$-matrix approximation relies on a more complex (four-point) effective interaction --- the so-called $T$ matrix --- yielding the following self-energy: 
\begin{equation}
\label{eq:SigGT}
	\Sigma^{GT}(1,2) = i \int G(4,3) T(1,3;2,4) d3 d4
\end{equation}
The natural idea of combining the ph and pp channels is also possible and has been explored, for example, in the Hubbard dimer within many-body perturbation theory (see Ref.~\onlinecite{Romaniello_2012} and references therein) and the uniform electron gas \cite{Loos_2016} within coupled-cluster theory. \cite{Shepherd_2014}

One of the key features of the $T$-matrix approximation is its exactness up to the second order thanks to the inclusion of second-order exchange diagrams.
This class of diagrams, which are particularly important in few-electron molecular systems \cite{Casida_1991,Ortiz_2013,Hirata_2015,Hirata_2017} and explain the improvement brought by the second-order screened exchange (SOSEX) correction applied to $GW$, \cite{Romaniello_2009a,Ren_2015,Loos_2018b} are well known to be missing in the $GW$ approximation. Moreover, unlike $W$ in the $GW$ approximation, the $T$-matrix
approximation also contains spin-flip terms; the spin structure of the $T$-matrix allows one to describe important processes like the emission of spin waves in ferromagnetics. \cite{Zhukov_2004}

In this work, we focus on neutral excitations and we explore how the $T$-matrix approximation performs within the Bethe-Salpeter equation (BSE) of many-body perturbation theory.\cite{Salpeter_1951,Strinati_1988,Blase_2018,Blase_2020}

Let us consider closed-shell electronic systems consisting of $N$ electrons and $K$ one-electron basis functions.
The number of singly-occupied and virtual (\ie, unoccupied) spinorbitals are $O = N$ and $V = K - O$, respectively.
Let us denote as $\SO{p}(\bx)$ the $p$th spinorbital and $\e{p}{}$ its one-electron energy.
The composite variable $\bx = (\sigma,\br)$ gathers spin ($\sigma$) and spatial ($\br$) variables.
We assume real quantities throughout this manuscript, $i$, $j$, $k$, and $l$ are occupied orbitals, $a$, $b$, $c$, and $d$ are unoccupied orbitals, $p$, $q$, $r$, and $s$ indicate arbitrary orbitals, $m$ labels single excitations, while $n$ labels double electron attachments or double electron detachments.

\section{Charged excitations}
\label{sec:charged}
By definition, \alert{in the quasiparticle approximation,} the one-body Green's function is \cite{Martin_2016}
\begin{equation}
\label{eq:G}
	G(\bx_1,\bx_2;\omega) 
	= \sum_i \frac{\SO{i}(\bx_1) \SO{i}(\bx_2)}{\omega - \e{i}{} - i\eta}	
	+ \sum_a \frac{\SO{a}(\bx_1) \SO{a}(\bx_2)}{\omega - \e{a}{} + i\eta}	
\end{equation}
where $\eta$ is a positive infinitesimal, and its nature is completely defined by the set of orbitals and corresponding energies that are used to build it.
For example, $G^{\HF}(\bx_1,\bx_2;\omega)$ is the Hartree-Fock (HF) Green's function built from the HF spinorbitals $\SO{p}^{\HF}(\bx)$ and energies $\e{p}{\HF}$.

Contrary to the $GW$ approximation which relies on the (two-point) dynamically-screened Coulomb potential $W$ computed from a ph-random-phase approximation (ph-RPA) problem to target charged excitations, \cite{Hedin_1965,Aryasetiawan_1998,Onida_2002,Martin_2016,Reining_2017,Golze_2019} here we consider the $GT$ approximation where one employs the (four-point) $T$ matrix obtained from solving the pp-RPA equations.

The non-Hermitian pp-RPA problem reads \cite{Schuck_Book,vanAggelen_2013,Peng_2013,Scuseria_2013,Yang_2013,Yang_2013b,vanAggelen_2014,Yang_2014a,Zhang_2015,Zhang_2016,Bannwarth_2020}
\begin{equation}
\label{eq:LR-RPA}
	\begin{pmatrix}
		\bA{}{\ppRPA}			&	\bB{}{\ppRPA}	\\
		-\T{(\bB{}{\ppRPA})}	&	-\bC{}{\ppRPA}	\\
	\end{pmatrix}
	\cdot
	\begin{pmatrix}
		\bX{n}{N\pm2}	\\
		\bY{n}{N\pm2}	\\
	\end{pmatrix}
	=
	\Om{n}{N\pm2}
	\begin{pmatrix}
		\bX{n}{N\pm2}	\\
		\bY{n}{N\pm2}	\\
	\end{pmatrix}
\end{equation}
where the elements of the various matrices are defined as
\begin{subequations}
\begin{align}
	A_{ab,cd}^{\ppRPA} & = \delta_{ab} \delta_{cd} (\e{a}{} + \e{b}{}) + \mel{ab}{}{cd}
	\\ 
	B_{ab,ij}^{\ppRPA} & = \mel{ab}{}{ij}
	\\ 
	C_{ij,kl}^{\ppRPA} & = - \delta_{ik} \delta_{jl} (\e{i}{} + \e{j}{}) +\mel{ij}{}{kl}
\end{align}
\end{subequations}
and 
\begin{equation}
	\mel{pq}{}{rs} = \braket{pq}{rs} - \braket{pq}{sr}
\end{equation}
are two-electron integrals in the spinorbital basis, \ie,
\begin{equation}
	\braket{pq}{rs} = \iint \SO{p}(\bx_1) \SO{q}(\bx_2) \frac{1}{\abs{\br_1 - \br_2}} \SO{r}(\bx_1) \SO{s}(\bx_2)  d\bx_1 d\bx_2
\end{equation}
The pp-RPA problem yields, in the absence of instabilities (which should not appear in Coulombic systems with repulsive interactions only \cite{Scuseria_2013}), $\Nvir(\Nvir-1)/2$ positive eigenvalues $\Om{n}{N+2}$ and $\Nocc(\Nocc-1)/2$ negative eigenvalues $\Om{n}{N-2}$, which  correspond respectively to double attachments and double detachments.
The pp-RPA correlation energy is given by \cite{Peng_2013,Scuseria_2013}
\begin{equation}
\begin{split}
	\Ec^\ppRPA 
	& = + \sum_n \Om{n}{N+2} - \Tr(\bA{}{\ppRPA}) 
	\\
	& = - \sum_n \Om{n}{N-2} - \Tr(\bC{}{\ppRPA})
\end{split}
\end{equation}

Considering the time structure of the $T$-matrix approximation as $T(1,3;2,4)=-\delta(t_1-t_3)\delta(t_2-t_4)\cT(\bx_1,\bx_2;\bx_4,\bx_3;t_1-t_2)$, \cite{Martin_2016} the  frequency-dependent $T$-matrix self-energy can be obtained from the Fourier transform of Eq.~\eqref{eq:SigGT} as 
\begin{multline}
    \Sigma^{GT}(\bx_1,\bx_2;\omega) 
    = -i \int d\bx_3 d\bx_4 \int\frac{d\omega'}{2\pi} G(\bx_4,\bx_3;\omega') \\
    \times \cT(\bx_1,\bx_3;\bx_2,\bx_4;\omega+\omega')
\end{multline}

The correlation part of $T$ matrix can be constructed from the knowledge of the pp-RPA eigenvalues and eigenvectors. 
In the spinorbital basis, it is defined as $\cT^c_{pq,rs}=\cT_{pq,rs}-\mel{pq}{}{rs}$ and it has the following form \cite{Zhang_2016}
\begin{equation}
\label{eq:T}
	\cT^c_{pq,rs}(\omega) 
		= \sum_{n} \frac{\braket*{pq}{\chi_n^{N+2}}\braket*{rs}{\chi_n^{N+2}}}{\omega - \Om{n}{N+2} + i\eta}
		- \sum_{n} \frac{\braket*{pq}{\chi_n^{N-2}}\braket*{rs}{\chi_n^{N-2}}}{\omega - \Om{n}{N-2} - i\eta}
\end{equation}
with
\begin{subequations}
\begin{align}
	\braket*{pq}{\chi_n^{N+2}} & = \sum_{c < d} \mel{pq}{}{cd} X_{cd,n}^{N+2} + \sum_{k < l}  \mel{pq}{}{kl} Y_{kl,n}^{N+2}
	\\
	\braket*{pq}{\chi_n^{N-2}} & = \sum_{c < d} \mel{pq}{}{cd} X_{cd,n}^{N-2} + \sum_{k < l}  \mel{pq}{}{kl} X_{kl,n}^{N-2}
\end{align}
\end{subequations}
Combining Eqs.~\eqref{eq:G} and \eqref{eq:T}, the correlation part of the $T$-matrix self-energy reads \cite{Romaniello_2012,Martin_2016,Zhang_2017,Li_2021b}
\begin{equation}
\label{eq:SigGTpq}
	\Sigma^{GT}_{pq}(\omega)
	= \sum_{in} \frac{\braket*{pi}{\chi_n^{N+2}}\braket*{qi}{\chi_n^{N+2}}}{\omega + \e{i}{} - \Om{n}{N+2} + i\eta}
	+ \sum_{an} \frac{\braket*{pa}{\chi_n^{N-2}}\braket*{qa}{\chi_n^{N-2}}}{\omega + \e{a}{} - \Om{n}{N-2} - i\eta}
\end{equation}
While the dynamical $GW$ self-energy corresponds to the downfolding of the 2h1p and 2p1h configurations on the 1h and 1p configurations via their coupling with the 1h1p configurations, respectively, \cite{Bintrim_2021a} Eq.~\eqref{eq:SigGTpq} shows that, in the case of the $T$-matrix approximation, the same 2h1p and 2p1h configurations are downfolded on the 1p and 1h configurations via their coupling with the 2h and 2p configurations, respectively.

Within the (perturbative) one-shot $GT$ scheme (labeled as $G_0T_0$ in the following), the quasiparticle energies are obtained via linearization of the quasiparticle equation, \cite{Strinati_1980,Hybertsen_1985a,Hybertsen_1986,Godby_1988,Linden_1988,Northrup_1991,Blase_1994,Rohlfing_1995,Shishkin_2007} \ie,
\begin{equation}
\label{eq:G0T0}
	\e{p}{G_0T_0} = \e{p}{\HF} + Z_p \Sigma^{GT}_{pp}(\e{p}{\HF})
\end{equation}
where we have assumed a HF starting point and
\begin{equation}
	Z_p = \qty[ 1 - \eval{\pdv{\Sigma^{T}_{pp}(\omega)}{\omega}}_{\omega = \e{p}{\HF}} ]^{-1}
\end{equation}
is the renormalization factor or weight of the quasiparticle solution.
Other levels of (partial) self-consistency can be considered like the \textit{``eigenvalue''} self-consistent $GT$ (ev$GT$)  \cite{Hybertsen_1986,Shishkin_2007,Blase_2011,Faber_2011,Rangel_2016,Gui_2018} or the quasiparticle self-consistent $GT$ (qs$GT$) \cite{Faleev_2004,vanSchilfgaarde_2006,Kotani_2007,Ke_2011,Kaplan_2016} schemes.

\section{Neutral excitations}
\label{sec:neutral}
Like the one-body Green's function is the pillar of the $GW$ and $GT$ approximations, the two-body Green's function $G_2$ is the central quantity of the BSE formalism of many-body perturbation theory \cite{Salpeter_1951,Strinati_1988,Blase_2018,Blase_2020} via its link with the two-body correlation function $L$ which satisfies the following Dyson equation 
\begin{multline}
	 L(1,2;1',2') = L_0(1,2;1',2') 
	 \\
	 + \int L_0(1,4;1',3) \Xi(3,5;4,6) L(6,2;5,2') d3d4d5d6
\end{multline}
where
\begin{subequations}
\begin{align}
\label{eq:L0}
	 iL_0(1,4;1',3) & = G(1,3) G(4,1')
	 \\
\label{eq:L}
	 iL(1,2;1',2') & = -G_2(1,2;1',2') + G(1,1')G(2,2')
\end{align}
\end{subequations}
and 
\begin{equation}
\label{eq:Xi}
	\Xi(3,5;4,6) = i \fdv{\Sigma(3,4)}{G(6,5)}
\end{equation}
is the so-called BSE kernel that takes into account the variation of $\Sigma$ with respect to the variation of $G$.
By taking into account the interaction of the excited electron and its hole left behind (the infamous excitonic effect), the BSE is able to faithfully model (neutral) optical excitations as measured by absorption spectroscopy. 
The moderate cost of the BSE [which scales as $\order*{K^4}$ in its standard implementation] and its all-round accuracy are the main reasons behind its growing popularity in the molecular electronic structure community. \cite{Rohlfing_1999a,Horst_1999,Puschnig_2002,Tiago_2003,Rocca_2010,Boulanger_2014,Jacquemin_2015a,Bruneval_2015,Jacquemin_2015b,Hirose_2015,Jacquemin_2017a,Jacquemin_2017b,Rangel_2017,Krause_2017,Gui_2018,Blase_2018,Liu_2020,Blase_2020,Holzer_2018a,Holzer_2018b,Loos_2020e,Loos_2021}

In order to target neutral (singly-)excited states, we first consider the static version of the BSE employing the $GT$ quasiparticle energies [see Eq.~\eqref{eq:G0T0}] as well as $T$-matrix kernel [\ie, $\Sigma^{\GT}$ in Eq.~\eqref{eq:Xi}].
In this case, the BSE@$GT$ linear eigenvalue problem simply reads
\begin{equation}
\label{eq:BSE}
	\begin{pmatrix}
		\bA{}{\BSE}		&	\bB{}{\BSE}	\\
		-\bB{}{\BSE}	&	-\bA{}{\BSE}	\\
	\end{pmatrix}
	\cdot
	\begin{pmatrix}
		\bX{m}{\BSE}	\\
		\bY{m}{\BSE}	\\
	\end{pmatrix}
	=
	\Om{m}{\BSE}
	\begin{pmatrix}
		\bX{m}{\BSE}	\\
		\bY{m}{\BSE}	\\
	\end{pmatrix}
\end{equation}
with 
\begin{subequations}
\begin{align}
	A_{ia,jb}^{\BSE} & = \delta_{ij} \delta_{ab} (\e{a}{\GT} - \e{i}{\GT}) + \mel{ib}{}{aj} + \cT^c_{ib,aj}(\omega=0)
	\\ 
	B_{ia,jb}^{\BSE} & = \mel{ij}{}{ab} + \cT^c_{ij,ab}(\omega=0)
\end{align}
\end{subequations}
The eigenvalues $\Om{m}{\BSE}$ of Eq.~\eqref{eq:BSE} provide $\Nocc\Nvir$ singlet (\ie, spin-conserved) and $\Nocc\Nvir$ triplet  (\ie, spin-flip) single excitations.
Note that the spin structure of the BSE@$GT$ equations is analogous to the BSE@$GW$ version, \cite{Monino_2021} and one can compute separately singlet and triplet excitation energies.
Neglecting the coupling between excitations and deexcitations, \ie, $\bB{}{\BSE} = \bO$, is known as the Tamm-Dancoff approximation (TDA).

Due to the frequency-independent nature of the static BSE, it is well known that one cannot access double (and higher) excitations.
\cite{Loos_2019,Romaniello_2009b,Sangalli_2011,Loos_2020h,Authier_2020,Monino_2021}
In order to go beyond the static approximation, it is possible to consider, within the dynamical TDA (dTDA) that neglects the frequency dependence of the coupling block $\bB{}{}$, the dynamical version of the BSE (dBSE). \cite{Strinati_1988,Romaniello_2009b,Loos_2020h}
In this case, one must solve the (non-linear) dynamical eigenvalue problem
\begin{equation}
\label{eq:dBSE}
	\begin{pmatrix}
		\bA{}{\dBSE}(\Om{S}{})	&	\bB{}{\BSE}	\\
		-\bB{}{\BSE}	&	-\bA{}{\dBSE}(-\Om{S}{})	\\
	\end{pmatrix}
	\cdot
	\begin{pmatrix}
		\bX{S}{\dBSE}	\\
		\bY{S}{\dBSE}	\\
	\end{pmatrix}
	=
	\Om{S}{}
	\begin{pmatrix}
		\bX{S}{\dBSE}	\\
		\bY{S}{\dBSE}	\\
	\end{pmatrix}
\end{equation}
with 
\begin{equation}
	A_{ia,jb}^{\dBSE}(\omega) = \delta_{ij} \delta_{ab} (\e{a}{\GT} - \e{i}{\GT}) + \mel{ib}{}{aj} + \widetilde{\cT}^c_{ib,aj}(\omega)
\end{equation}
where, following Strinati's seminal work, \cite{Strinati_1988} one can derive the following expression for the elements of the dynamical $T$ matrix
\begin{equation}
\label{eq:dT}
\begin{split}
	\widetilde{\cT}^c_{ib,aj}(\omega) 
		& = \sum_{n} \frac{\braket*{ib}{\chi_n^{N+2}}\braket*{aj}{\chi_n^{N+2}}}{\omega - \Om{n}{N+2} + (\e{i}{\GT} + \e{j}{\GT}) + i\eta}
		\\
		& + \sum_{n} \frac{\braket*{ib}{\chi_n^{N-2}}\braket*{aj}{\chi_n^{N-2}}}{\omega + \Om{n}{N-2} - (\e{a}{\GT} + \e{b}{\GT}) + i\eta}
\end{split}
\end{equation}
from which, one can check that we recover the static expression \eqref{eq:T} in the limit $\Om{n}{N\pm2} \to \infty$.
Equation \eqref{eq:dT} highlights the interesting dynamical structure of the $T$ matrix, where, similarly to the dBSE@$GW$ scheme, \cite{Strinati_1988,Romaniello_2009b,Loos_2020h} the 2h2p configurations are downfolded on the 1h1p configurations. \cite{Bintrim_2021b}
Additional details about the derivation of Eq.~\eqref{eq:dT} are reported in Appendix \ref{app:dBSE}.

Because solving a non-linear eigenvalue problem is computationally challenging, here we rely on the perturbative scheme developed in Ref.~\onlinecite{Loos_2020h} in order to access \textit{dynamically-corrected} single excitations for which additional relaxation effects coming from higher excitations are taken into account.
\cite{Rohlfing_2000,Romaniello_2009b,Ma_2009a,Ma_2009b,Zhang_2013,Rebolini_2016,Olevano_2019,Loos_2020h,Authier_2020,Monino_2021}
Below, we quickly recap this dynamical perturbative scheme.

Based on Rayleigh-Schr\"odinger perturbation theory, the non-linear eigenproblem \eqref{eq:dBSE} can be split as a zeroth-order static reference and a first-order dynamic perturbation, such that
\begin{multline}
\label{eq:LR-PT}
	\begin{pmatrix}
		\bA{}{\dBSE}(\Om{S}{})		&	\bB{}{\BSE}	\\
		-\bB{}{\BSE}	&	-\bA{}{\BSE}(-\Om{S}{})	\\
	\end{pmatrix}
	\\
	=
	\begin{pmatrix}
		\bA{}{\BSE}	&	\bB{}{\BSE}	
		\\
		-\bB{}{\BSE}	&	-\bA{}{\BSE}	
		\\
	\end{pmatrix}
	+
	\begin{pmatrix}
		\bA{}{(1)}(\Om{S}{})		&	\bO	\\
		\bO	&	-\bA{}{(1)}(-\Om{S}{})	\\
	\end{pmatrix}
\end{multline}
with
\begin{equation}
	\label{eq:BSE-A1}
	A_{ia,jb}^{(1)}(\omega) = \widetilde{\cT}^c_{ib,aj}(\omega) - \cT^c_{ib,aj}(\omega = 0)
\end{equation}
As usual, one can naturally expand the $S$th BSE excitation energy and its corresponding eigenvector as
\begin{subequations}
\begin{gather}
	\Om{S}{} = \Om{S}{\BSE} + \Om{S}{(1)} + \ldots,
	\\
	\begin{pmatrix}
		\bX{S}{}	\\
		\bY{S}{}	\\
	\end{pmatrix}
	= 
	\begin{pmatrix}
		\bX{S}{\BSE}	\\
		\bY{S}{\BSE}	\\
	\end{pmatrix}
	+
	\begin{pmatrix}
		\bX{S}{(1)}	\\
		\bY{S}{(1)}	\\
	\end{pmatrix}
	+ \ldots
\end{gather}
\end{subequations}
Solving the static BSE [see Eq.~\eqref{eq:BSE}] yields the (zeroth-order) static $\Om{S}{\BSE}$ excitation energies and their corresponding eigenvectors $\bX{S}{\BSE}$ and $\bY{S}{\BSE}$. 
The first-order correction to the $S$th excitation energy is, within the dTDA,
\begin{equation}
\label{eq:Om1-TDA}
	\Om{S}{(1)} = \T{(\bX{S}{\BSE})} \cdot \bA{}{(1)}(\Om{S}{\BSE}) \cdot \bX{S}{\BSE}
\end{equation}
This correction can be renormalized by computing, at no extra cost, the renormalization factor which reads
\begin{equation}
\label{eq:Z}
	\zeta_{S} = \qty[ 1 - \T{(\bX{S}{\BSE})} \cdot \eval{\pdv{\bA{}{(1)}(\Om{S}{})}{\Om{S}{}}}_{\Om{S}{} = \Om{S}{\BSE}} \cdot \bX{S}{\BSE} ]^{-1}
\end{equation}
This yields our final expression for the dynamically-corrected BSE excitation energies:
\begin{equation}
	\Om{S}{\text{dyn}} = \Om{S}{\text{stat}} + \Delta\Om{S}{\text{dyn}} = \Om{S}{\BSE} + \zeta_{S} \Om{S}{(1)}
\end{equation}
Note again that the present perturbative scheme does not allow to access double excitations as only excitations calculated within the static approach can be dynamically corrected.

\section{Computational details}
\label{sec:compdet}
The present formalism has been implemented in the electronic structure package \texttt{QuAcK} \cite{QuAcK} which is freely available at \url{https://github.com/pfloos/QuAcK}. 
We consider here only systems with closed-shell singlet ground states. 
Thus, the $GW$ and $GT$ calculations are performed by considering a (restricted) HF starting point and standard gaussian basis sets (defined with cartesian functions) are employed.
Note that all quasiparticle energies which are obtained via Eq.~\eqref{eq:G0T0} are corrected in the same way.
Finally, the infinitesimal $\eta$ is set to zero for all calculations.
\alert{The ev$GT$ and qs$GT$ schemes have been also implemented but are not considered here, mainly because, for the small molecular systems studied here (see below), we have observed very small differences between one-shot and self-consistent quasiparticle energies.}
Although the dynamical correction is computed in the dTDA throughout, the zeroth-order excitonic Hamiltonian [see Eq.~\eqref{eq:BSE}] is always the ``full'' BSE static Hamiltonian, i.e., without TDA. 
Reference full configuration interaction (FCI) calculations have been performed with {\QP}. \cite{Garniron_2019}

In terms of computational cost, the overall scaling of BSE@$GT$ is equivalent to BSE@$GW$ as they both correspond to seeking the lowest eigenvalues of a matrix of size $(2\Nocc\Nvir \times 2\Nocc\Nvir)$. 
Searching iteratively for the lowest eigenstates can be routinely performed via Davidson's algorithm with a $\order*{\Norb^4}$ computational cost. \cite{Davidson_1975}
The cost of the dynamical correction, which is thoroughly discussed in Ref.~\onlinecite{Loos_2020h}, is more expensive but is again equivalent in both formalisms.
The computational cost associated with the computation of the $T$-matrix and the screening $W$ both scale as $\order*{\Norb^6}$ in their standard implementation as one must obtain all the eigenvalues and eigenvectors of the pp-RPA and ph-RPA problems, respectively. \cite{Shenvi_2014} 
However, the prefactor of the pp-RPA calculation is significantly larger than its ph-RPA counterpart due to the larger size of the pp-RPA matrices and its non-Hermitian nature. \cite{Peng_2013,Scuseria_2013,Yang_2013,Yang_2013b,Yang_2014a,Zhang_2015,Zhang_2016}
Moreover, within the $T$-matrix formalism, one must compute both the singlet and triplet contributions of the $T$-matrix, while for singlet states, only the singlet part of $W$ is required.
Although similar approaches remain to be developed for the $T$-matrix formalism, contour deformation and density fitting techniques can be efficiently implemented in the case of $GW$ to reduce the scaling to $\order*{\Norb^3}$. \cite{Duchemin_2019,Duchemin_2020,Duchemin_2021}

\section{Results and discussion}
\label{sec:res}

\subsection{Excited states of the hydrogen molecule}
\label{sec:h2}

\begin{figure*}
	\includegraphics[width=\textwidth]{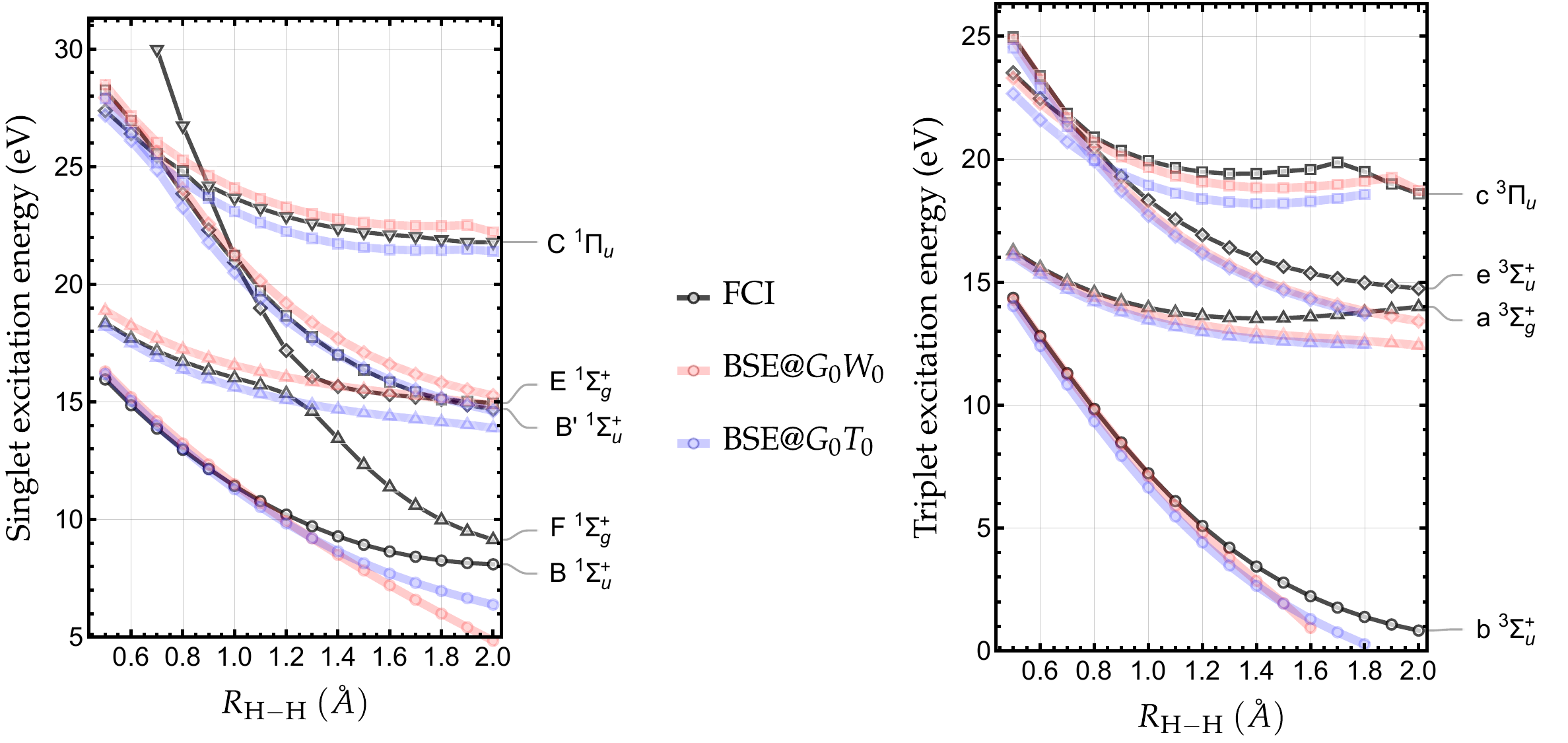}
	\caption{Singlet (left) and triplet (right) excitation energies (in \si{\eV}) of \ce{H2} as a function of the internuclear distance $R_{\ce{H-H}}$ (in \si{\angstrom}) computed at the FCI (black), BSE@$G_0W_0$ (blue), and BSE@$G_0T_0$ (red) levels with the cc-pVTZ basis.
	Raw data are reported in {\SupMat}.}
	\label{fig:H2}
\end{figure*}

\begin{figure}
	\includegraphics[width=\linewidth]{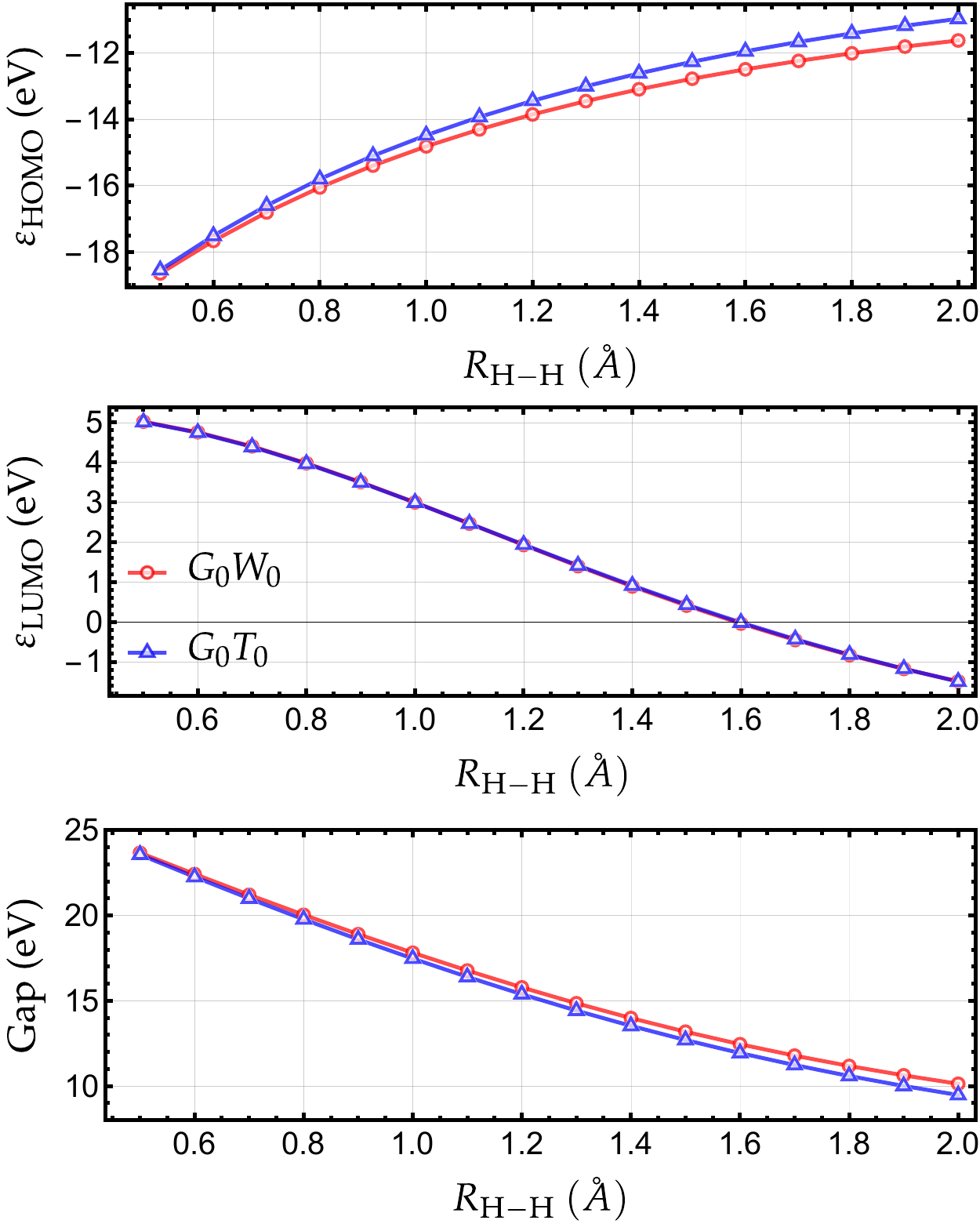}
	\caption{HOMO and LUMO quasiparticle energies as well as HOMO-LUMO gap (in \si{\eV}) of \ce{H2} as a function of the internuclear distance $R_{\ce{H-H}}$ (in \si{\angstrom}) computed at the $G_0W_0$ (blue) and $G_0T_0$ (red) levels with the cc-pVTZ basis.
	Raw data are reported in {\SupMat}.}
	\label{fig:H2_gap}
\end{figure}

\begin{figure*}
	\includegraphics[width=0.4\linewidth]{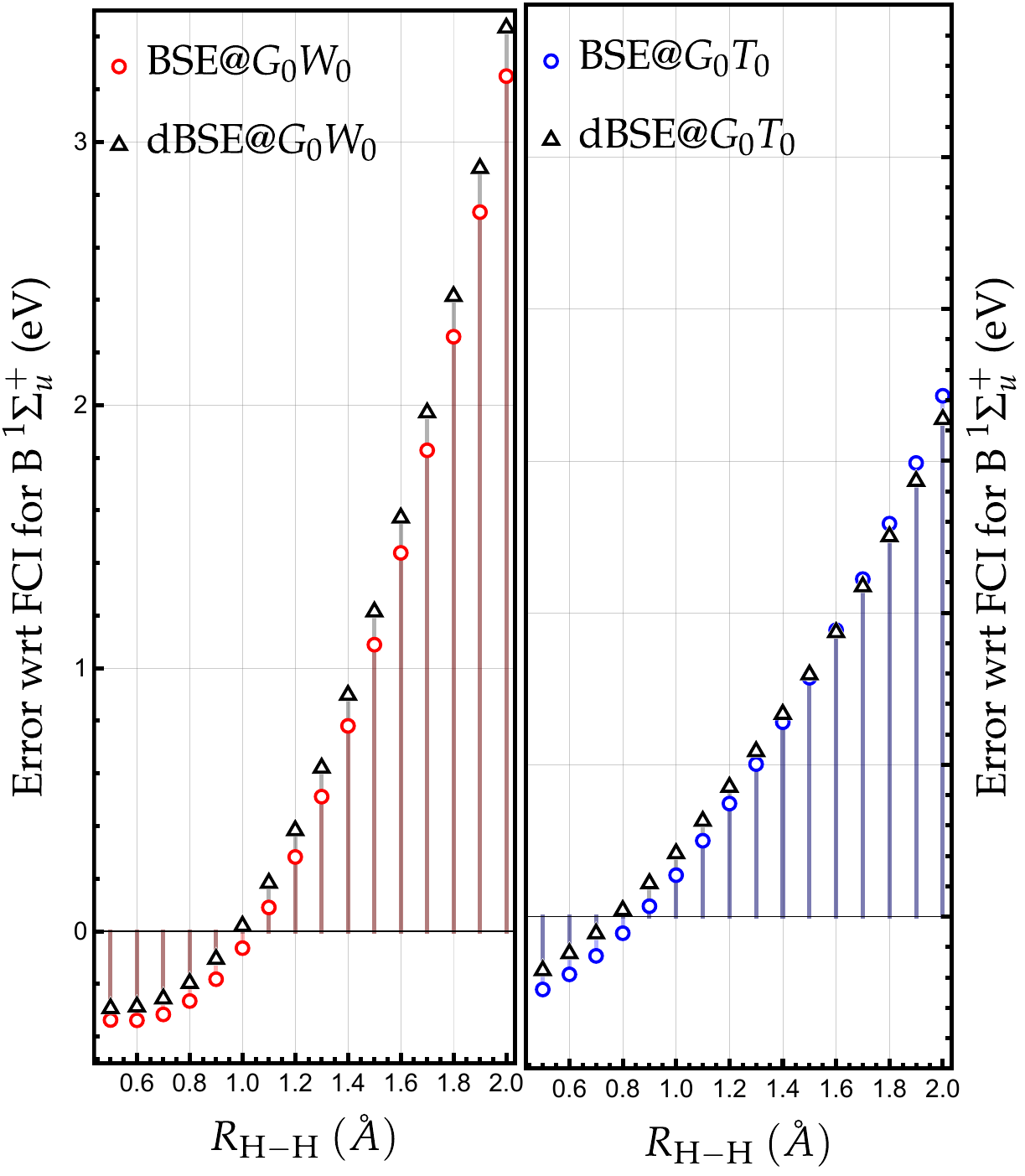}
	\hspace{0.1\linewidth}
	\includegraphics[width=0.4\linewidth]{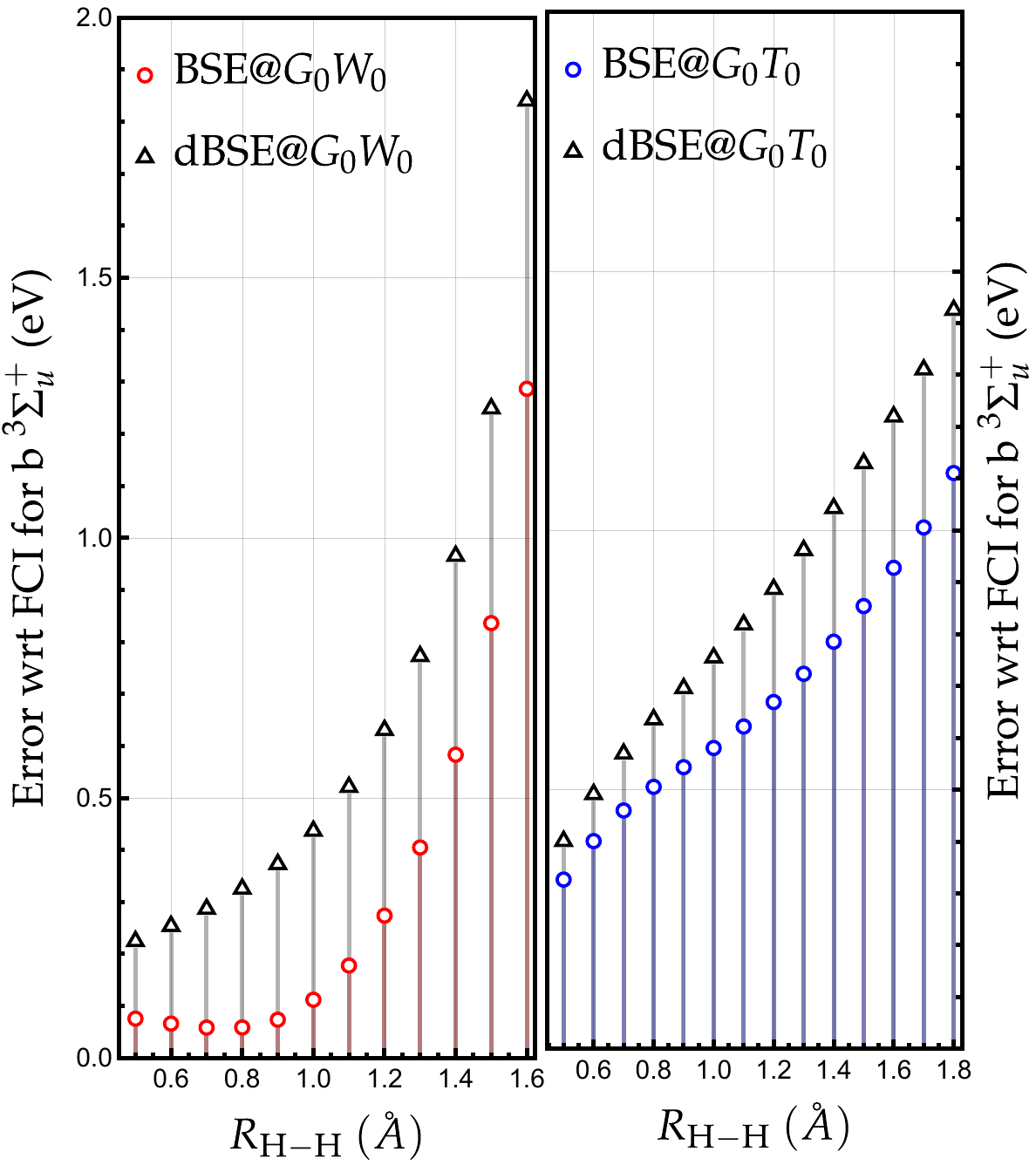}
	\caption{
	Error with respect to FCI for the lowest singlet (left) and the lowest triplet (right) excitation energies of \ce{H2} as a function of the internuclear distance $R_{\ce{H-H}}$ (in \si{\angstrom}) computed within the static schemes (BSE@$G_0W_0$ and BSE@$G_0T_0$) and the dynamically-corrected schemes (dBSE@$G_0W_0$ and dBSE@$G_0T_0$) .
	The cc-pVTZ basis is employed for all calculations.
	Raw data are reported in {\SupMat}.}
	\label{fig:H2_dBSE}
\end{figure*}

As a first didactical example, we consider the lowest singlet and triplet excited states of the hydrogen molecule \ce{H2} and the variation of their respective vertical transition energies upon dissociation.
The excitation energies associated with these low-lying excited states are represented in Fig.~\ref{fig:H2} as a function of the internuclear distance $R_{\ce{H-H}}$ at the FCI (black), BSE@$G_0W_0$ (blue), and BSE@$G_0T_0$ (red) levels with the cc-pVTZ basis.
The variation of the HOMO and LUMO quasiparticle energies as well as HOMO-LUMO gap computed at the $G_0W_0$ and $G_0T_0$ levels is depicted in Fig.~\ref{fig:H2_gap}.
This shows that, as already observed in the Hubbard dimer \cite{Romaniello_2012} and in molecular systems, \cite{Zhang_2017} the $G_0W_0$ and $G_0T_0$ quasiparticle energies are similar near the Fermi level.

Overall, as evidenced by Fig.~\ref{fig:H2}, the performances of BSE@$G_0W_0$ and BSE@$G_0T_0$ are analogous for this system.
For the lowest singlet excited state of $\text{B}\,{}^1\Sigma_u^+$ symmetry, the $T$-matrix-based formalism is slightly better when $R_{\ce{H-H}}$ increases but fails ultimately to reproduce the FCI results.
For the $\text{E}\,{}^1\Sigma_g^+$ state, BSE@$G_0T_0$ is more accurate than BSE@$G_0W_0$ for small bond length and the scenario is reversed after the avoided crossing with the doubly-excited state of $\text{F}\,{}^1\Sigma_g^+$ symmetry.
Of course, both formalisms cannot ``see'' the $\text{F}\,{}^1\Sigma_g^+$ states as the static BSE formalism is blind to double excitations.
Therefore, it cannot model properly the avoided crossing between $\text{E}\,{}^1\Sigma_g^+$ and $\text{F}\,{}^1\Sigma_g^+$ states.
For the $\text{B'}\,{}^1\Sigma_u^+$ and $\text{C}\,{}^1\Pi_u$ states, BSE@$G_0W_0$ and BSE@$G_0T_0$ reproduces fairly well the FCI potential energy curves with a modest preference for the latter.

Similar observations can be made for the triplet states, the $GW$- and $GT$-based formalisms yielding very similar excitation energies, except for the $\text{c}\,{}^3\Pi_u$ state for which BSE@$G_0W_0$ has clearly the edge.
Moreover, triplet instabilities seems to affect BSE@$G_0T_0$ slightly earlier than BSE@$G_0W_0$.

In Fig.~\ref{fig:H2_dBSE}, we show the energy shift provided by the dynamical correction for the lowest singlet and lowest triplet excited states of \ce{H2} as a function of $R_{\ce{H-H}}$. 
These dynamically-corrected schemes are labeled dBSE@$G_0W_0$ and dBSE@$G_0T_0$.
For the singlet state of $\text{B}\,{}^1\Sigma_u^+$ symmetry, the dynamical correction improves slightly the excitation energies at small internuclear distances for both schemes, while, for larger bond lengths, an improvement is only visible at the $T$-matrix level.
Note that, for this system with few electrons, the dynamical corrections are quite small in magnitude.
In the case of the triplet state of $\text{b}\,{}^3\Sigma_u^+$ symmetry, the dynamical correction worsens the results compared to FCI, especially in the case of BSE@$G_0W_0$.

\subsection{Excited states of beryllium hydride}
\label{sec:beh2}

\begin{figure*}
	\includegraphics[width=\textwidth]{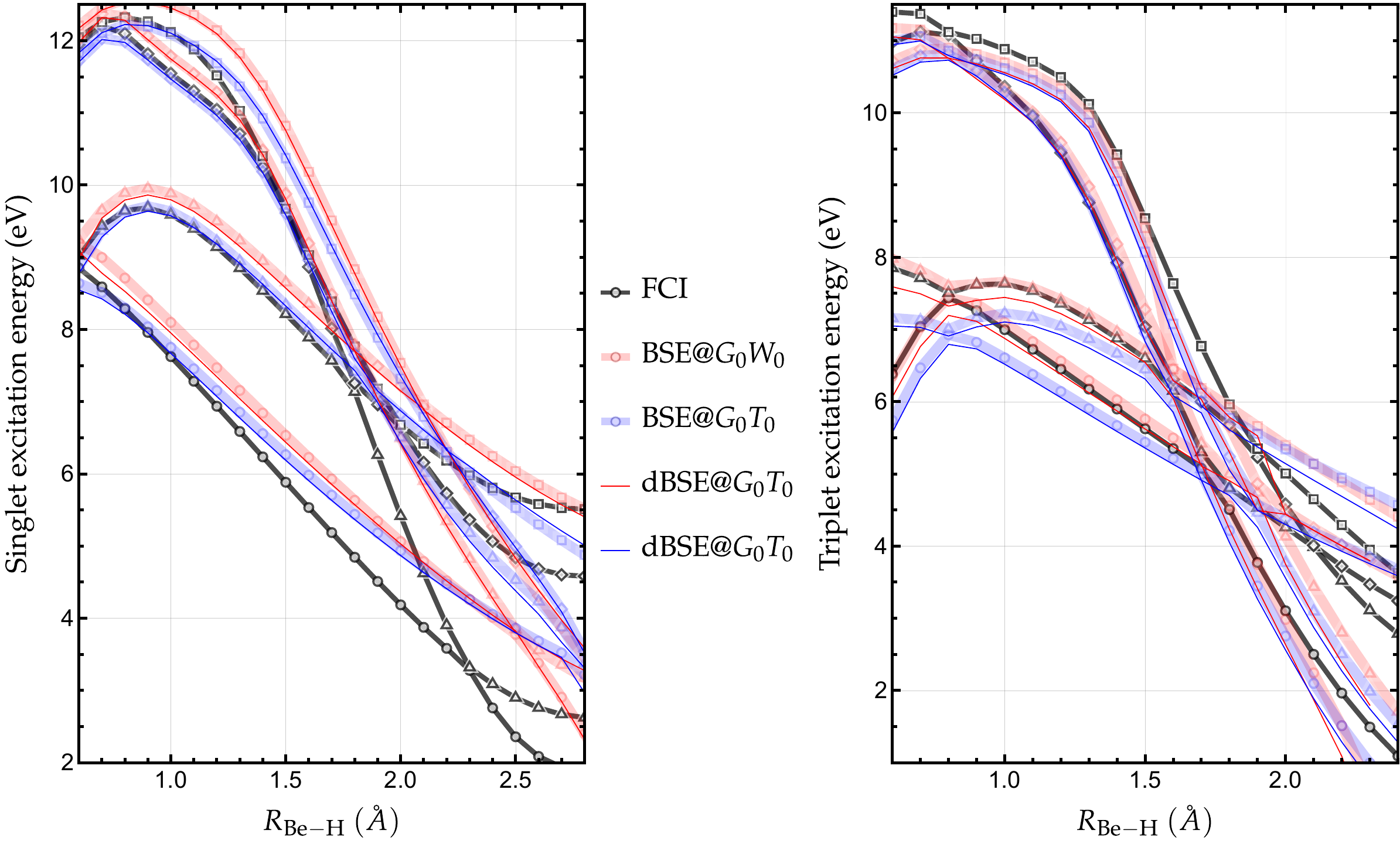}
	\caption{Singlet (left) and triplet (right) excitation energies (in \si{\eV}) of \ce{BeH2} as a function of the distance $R_{\ce{Be-H}}$ (in \si{\angstrom}) computed at the FCI (black), BSE@$G_0W_0$ (blue), and BSE@$G_0T_0$ (red) levels with the cc-pVDZ basis.
	The dynamically-corrected BSE excitation energies are represented as thin lines for dBSE@$G_0W_0$ (blue) and dBSE@$G_0T_0$ (red).
	Raw data are reported in {\SupMat}.}
	\label{fig:BeH2}
\end{figure*}

\begin{figure}
	\includegraphics[width=\linewidth]{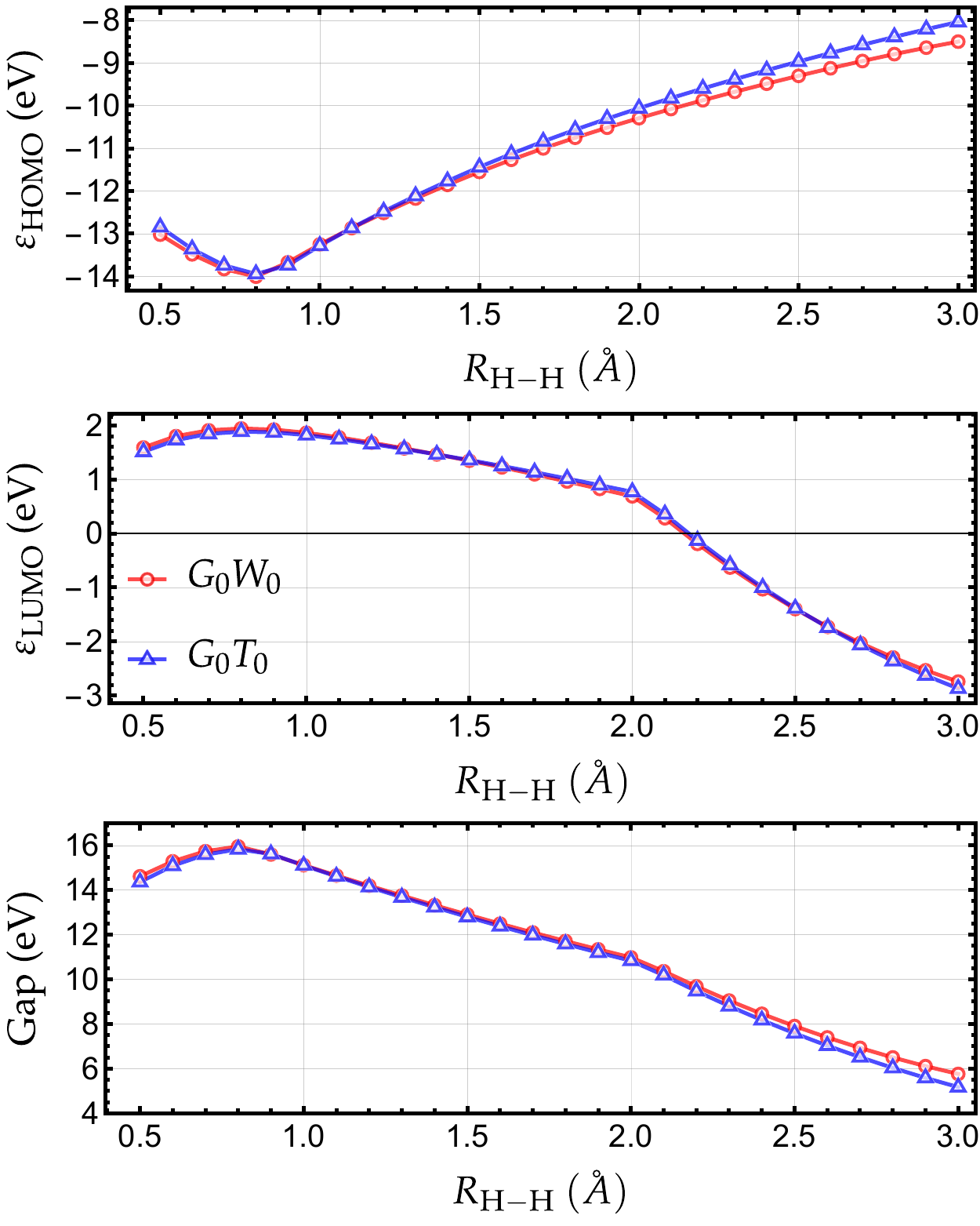}
	\caption{HOMO and LUMO quasiparticle energies as well as HOMO-LUMO gap (in \si{\eV}) of \ce{BeH2} as a function of the internuclear distance $R_{\ce{Be-H}}$ (in \si{\angstrom}) computed at the BSE@$G_0W_0$ (blue) and BSE@$G_0T_0$ (red) levels with the cc-pVDZ basis.
	Raw data are reported in {\SupMat}.}
	\label{fig:BeH2_gap}
\end{figure}

As a second example, we consider the symmetric dissociation of the linear molecule beryllium hydride (\ce{BeH2}), a system for which one can assume that the screening plays a more important role than in the previous example.
The variation of the lowest singlet and triplet excitation energies as a function of the distance $R_{\ce{Be-H}}$ is shown in Fig.~\ref{fig:BeH2}, while the quasiparticle energies of the frontier orbitals and the associated (fundamental) gap computed at the $G_0W_0$ and $G_0T_0$ levels is depicted in Fig.~\ref{fig:BeH2_gap}.
All calculations are performed with the cc-pVDZ basis.
Again, one notes that the $G_0W_0$ and $G_0T_0$ quasiparticle energies are very similar near the Fermi level.
Therefore, one can safely assume that any significant variation of the excitation energies computed within the $GW$- and $GT$-based formalisms originates mainly from their distinct kernel.
The excitation energies computed with the dynamical schemes, dBSE@$G_0W_0$ and dBSE@$G_0T_0$, are reported as thin solid lines.
Here, one can show that dynamical corrections improves in most cases the agreement between BSE and FCI.

For the four lowest singlet excited states (left panel of Fig.~\ref{fig:BeH2}), dBSE@$G_0T_0$ is clearly better than dBSE@$G_0W_0$, while the opposite trend is observed for the four lowest triplet states (right panel of Fig.~\ref{fig:BeH2}).
Note that, for large $R_{\ce{Be-H}}$, the two BSE-based schemes provide only a qualitative description of the excited states with errors of several \si{\eV}.
Nonetheless, the overall ordering of the excited states are globally respected.

\subsection{Excited states of water}
\label{sec:H2O}

\begin{figure*}
	\includegraphics[width=\linewidth]{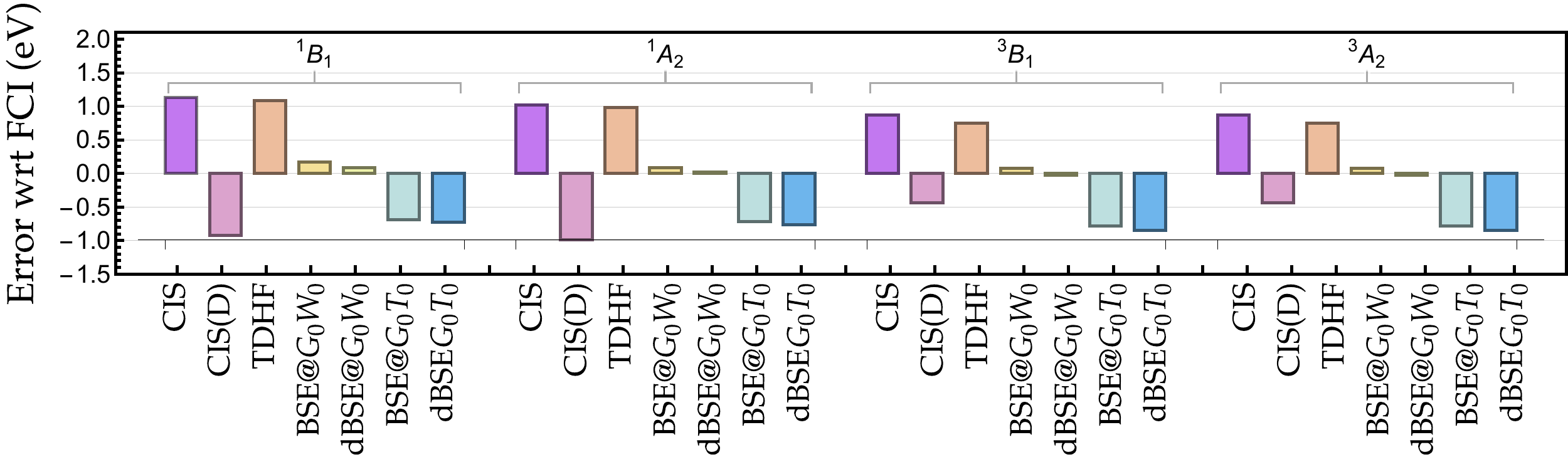}
	\caption{Error with respect to FCI for the singlet (left) and triplet (right) excitation energies (in \si{\eV}) of \ce{H2O} at equilibrium geometry computed at various levels of theory with the aug-cc-pVDZ basis.
	The geometry and the reference FCI values have been extracted from the QUEST database. \cite{Veril_2021}
	Raw data are reported in {\SupMat}.}
	\label{fig:H2O}
\end{figure*}

As a third and final example, we compute the excitation energies associated with the two lowest singlet and two lowest triplet excited states of water at equilibrium geometry (see Fig.~\ref{fig:H2O}).
Note that all these excited states are of Rydberg nature and correspond to $n \to 3s$ and $n \to 3p$ transitions for the $B_1$ and $A_2$ states, respectively. \cite{Loos_2018a}
In addition to the BSE-based models studied in the present manuscript, we have selected well-known wave function methods, \cite{Bene_1971,Head-Gordon_1994,Head-Gordon_1995,Dreuw_2005} namely CIS, CIS(D), TDHF, and FCI (taken as reference) and computed the excitation energies of these transitions.
It is worth mentioning here that the TDHF (or RPAx) equations within the TDA are strictly equivalent to the CIS equation, \cite{Dreuw_2005} and that CIS(D) is a simple perturbative doubles correction to CIS, and can be considered as an excited-state analog of second-order M{\o}ller-Plesset perturbation theory. \cite{Head-Gordon_1994,Head-Gordon_1995}

Two key observations can be made:
i) BSE@$G_0W_0$ is by far the best performer with a slight overestimation of the order of \SI{0.1}{eV} (as compared to FCI);
ii) BSE@$G_0T_0$ systematically underestimates the excitation energies [similarly to CIS(D)] and outperforms CIS, CIS(D) and TDHF for the singlet states only. 
These general trends are also observed for other systems and they nicely evidence the crucial role of the screening in $GW$, hinting that a screened version of the $T$-matrix formalism as proposed in Ref.~\onlinecite{Romaniello_2012} might be a promising way for improvement.

\section{Conclusion}
\label{sec:ccl}
We have derived and implemented, for the first time, the static and dynamic Bethe-Salpeter equations when one considers $T$-matrix quasiparticle energies as well as a $T$-matrix-based kernel.
The performance of the static scheme and its perturbative dynamical correction have been assessed by computing the neutral excited states of several molecular systems.
Our results suggest that, in the context of the computation of molecular excitation energies, the BSE@$GT$ formalism performs best in few-electron systems where the electron density remains low.
\alert{The overall accuracy of the present scheme still needs to be assessed for larger systems (where the screening is known to be more important). 
For such purposes, a comprehensive benchmark study would be required and we are planning to do so in the future.}

It would be interesting to investigate its performance for the computation of ground-state correlation energies within the adiabatic connection fluctuation dissipation formalism where BSE@$GW$ has been shown to be particularly outstanding. \cite{Maggio_2016,Holzer_2018b,Loos_2020e}
The combination of $GT$ and $GW$ via the range separation of the Coulomb operator to avoid double counting of the low-order diagrams is also a promising avenue.
Work along these lines are currently under progress.
Finally, the unrestricted and spin-flip extensions of the present formalism are currently being developed.

\begin{acknowledgements}
The authors thank Roberto Orlando for useful discussions.
PFL thanks the European Research Council (ERC) under the European Union's Horizon 2020 research and innovation programme (Grant agreement No.~863481) for funding.
This project has also received financial support from the CNRS through the 80|Prime program and has been supported through the EUR grant NanoX ANR-17-EURE-0009 in the framework of the ``Programme des Investissements d’Avenir''.
\end{acknowledgements}

\onecolumngrid
\appendix
\section{BSE with a dynamical $T$-matrix kernel}
\label{app:dBSE}

In order to derive the dynamical kernel $\widetilde{\cT}^c_{ib,aj}$ given in Eq.~\eqref{eq:dT}, we follow Ref.~\onlinecite{Strinati_1980} (see also Ref.~\onlinecite{Loos_2020e}) and start from the equation for the BSE amplitude
\begin{equation}
\label{eq:Strinati_1}
	\chi_S(1,1') = \int d3d4d5d6L_0(1,4,1',3)\Xi(3,5,4,6) \chi_S(6,5)
\end{equation}
where $L_0$ is given by Eq.~\eqref{eq:L0} and the $T$-matrix kernel is
\begin{equation}
\label{eq:T-matrix_kernel}
	\Xi(3,5,4,6) =  i\frac{\delta \Sigma(3,4)}{\delta G(6,5)}\approx -T(3,5,4,6)
\end{equation}
Equation (\ref{eq:Strinati_1}) is derived by assuming that i) the (resonant) pole $\omega_S=E_S-E_0>0$ of $L$ is isolated from the other poles (which is usually the case for neutral excitations in finite systems), and  ii) the poles of $L_0$ are different from $\omega_S$ (which is also generally the case).
The so-called $T$-matrix self-energy $\Sigma$ is given by Eq.~\eqref{eq:SigGT} with \cite{Martin_2016,Romaniello_2012}
\begin{equation}
\label{eq:T-matrix}
	T(3,8,4,7) =
	- v(3,8)\delta(3,4)\delta(7,8)
	+ v(3,8)\delta(4,8)\delta(3,7)
	+ i \int d1'd2' v(3,8)G(3,1')G(8,2')T(1',2',4,7)
\end{equation}  
where $v$ is the bare Coulomb operator and we neglect the functional derivative $\delta T/\delta G$ in the kernel $\Xi$. 
The first two terms in the right-hand side of Eq.~\eqref{eq:T-matrix} are the Hartree and exchange contributions to the $T$-matrix, whereas the last term is the correlation contribution.
Making the time dependence of Eq.~\eqref{eq:Strinati_1} explicit and defining $T(3,5,4,6)=- \delta(\tau^+_{35})\delta(\tau^+_{64})\cT(\bx_3,\bx_5,\bx_4,\bx_6;\tau_{34})$, one gets 
\begin{equation}
\label{eq:DYN_T}
	\chi_S(\bx_1,\bx_{1'},\tau_{11'})e^{-i\omega_S(t_1+t_{1'})/2}
	=-i\int d\bx_3d\bx_4d\bx_5d\bx_6\int dt_3dt_4 G(\bx_1,\bx_3;\tau_{13})G(\bx_4,\bx_{1'};\tau_{11'})
	\cT(\bx_3,\bx_5,\bx_4,\bx_6;\tau_{34})\chi_S(\bx_6,\bx_5;-\tau_{34})e^{-i\omega_S\tau_{34}/2}
\end{equation}
[where $\tau_{ij}=t_i-t_j$ and $\tau^+_{ij}=t^+_i-t_j$ with $t^+_i=t_i+\eta$ ($\eta \to 0^+$)] and
\begin{multline}
\label{eq:T_bar}
	\cT(\bx_3,\bx_5,\bx_4,\bx_6;\tau_{34}) 
	= v(\bx_3,\bx_5)\delta(\tau_{34})\delta(\bx_3,\bx_4)\delta(\bx_6,\bx_5)-v(\bx_3,\bx_5)\delta(\tau_{34})	\delta(\bx_3,\bx_6)\delta(\bx_4,\bx_5)
	\\
	+ i \int d\bx_7 d\bx_8\int dt_ 7 v(\bx_3,\bx_5)G(\bx_3,\bx_7;\tau_{37})G(\bx_5,\bx_8;\tau^+_{37})\cT(\bx_7,\bx_8,\bx_4,\bx_6;\tau_{74})
\end{multline}
Using the Fourier transform $G(\tau)=\int \frac{d\omega}{2\pi}G(\omega)e^{-i\omega\tau}$, changing variable from $t_3$ to $\tau_{34}$ and taking the limit $t_{1'}=t_1^+$, we have
\begin{multline}
	\chi_S(\bx_1,\bx_{1'},0^-)
	= -i\int d\bx_3 d\bx_4 d\bx_5 d\bx_6\int d\tau_{34} \int \frac{d\omega'}{2\pi}G(\bx_1,\bx_3;\omega'+\omega_S)G(\bx_4,\bx_{1'};\omega')e^{i\omega'\tau_{34}}
	\\
	 \times \cT(\bx_3,\bx_5,\bx_4,\bx_6;\tau_{34})\chi_S(\bx_6,\bx_5;-\tau_{34})e^{-i\omega_S\tau_{34}/2}
\end{multline}
Using the Lehman representation of the one-body Green's function in the quasiparticle approximation given by Eq.~\eqref{eq:G}, and multiply the left- and right-hand sides by $(\e{a}{}-\e{i}{}-\omega_S)\int d\bx_1 d\bx^{\prime}_1 \SO{a}(\bx_1)\SO{i}(\bx_1')$, we obtain  
\begin{multline}
\label{eq:General_DYN}
	\qty(\e{a}{}-\e{i}{}-\omega_S)\int d\bx_1d\bx^{\prime}_1 \SO{a}(\bx_1)\SO{i}(\bx_1')  \chi_S(\bx_1,\bx_{1'},0^-) =
	-\int d\bx_3 d\bx_4 d\bx_5 d\bx_6\int d\tau_{34}\SO{a}(\bx_3)\SO{i}(\bx_4)
	\\
	\times \qty[\Theta(\tau_{34})e^{i\e{i}{}\tau_{34}}+\Theta(-\tau_{34})e^{i(\e{a}{}-\omega_S)\tau_{34}} ]
	\cT(\bx_3,\bx_5,\bx_4,\bx_6;\tau_{34})\chi_S(\bx_6,\bx_5;-\tau_{34})e^{i\omega_S\tau_{34}/2}
\end{multline}
(where $\Theta$ is the Heaviside step function) using the fact that
\begin{equation}
	\Theta(\pm\tau)e^{-i\alpha \tau}=\mp\frac{1}{2\pi i}\lim_{\eta\to 0^+}\int d\omega \frac{1}{\omega-\alpha\pm i\eta}e^{-i\omega\tau}
\end{equation}

For the resonant case, \ie, $\omega_S>0$, we have
\begin{equation}
\label{eq:chi_tau}
	\chi_S(\bx_1,\bx_{1'},\tau_1) 
	=-e^{i\omega_S\abs{\tau_1}/2} \sum_{jb}\SO{b}(\bx_1)\SO{j}(\bx_{1'}) \mel{N}{\hat{c}^\dagger_{j}\hat{c}_b}{N,S}
  	\qty[\Theta(\tau_1)e^{-i\e{b}{}\tau_1}+\Theta(-\tau_1)e^{-i\e{j}{}\tau_1} ]
\end{equation}
  where $\hat{c}^\dagger_{p}$ and $\hat{c}_p$ are the usual creation and annihilation operators, respectively, and $\ket{N}$ and $\ket{N,S}$ are the ground state and the $S$th excited state, respectively, of the $N$-electron system.
After some algebraic steps, one gets
\begin{multline}
	-\qty(\e{a}{}-\e{i}{}-\omega_S)\mel{N}{\hat{c}^\dagger_{i}\hat{c}_a}{N,S}
	\\
    = \sum_{jb}\mel{N}{\hat{c}^\dagger_{j}\hat{c}_b}{N,S}
    \qty{ \frac{i}{2\pi}\int d\omega \lim_{\eta \to 0^+}\ \cT_{ib,aj}(\omega) e^{-i\omega\eta} \qty[\frac{1}{\omega_S-\omega+\e{j}{}+\e{i}{}+i\eta}+\frac{1}{\omega_S+\omega-\e{b}{}-\e{a}{}+i\eta} ] }
\end{multline}
where we have defined 
\begin{equation}
	\cT_{ib,aj}(\tau_{34})=\int d\bx_3 d\bx_4 d\bx_5 d\bx_6 \SO{a}(\bx_3) \SO{i}(\bx_4)\cT(\bx_3,\bx_5,\bx_4,\bx_6;\tau_{34}) \SO{b}(\bx_6)\SO{j}(\bx_5).
\end{equation}
Using the definition $X_{ia,S}=\mel{N}{\hat{c}^\dagger_{i}\hat{c}_a}{N,S}$, we arrive at

\begin{equation}
	\label{eq:egval}
	\qty(\e{a}{}-\e{i}{}-\omega_S) X_{ia,S}+\sum_{jb}X_{jb,S}\mel{ib}{}{aj} + \sum_{jb}X_{jb,S}\widetilde{\cT}^c_{ib,aj}(\omega_S)=0
\end{equation}
where the spectral representation of the dynamical $T$-matrix is
\begin{equation}
	\widetilde{\cT}^c_{ib,aj}(\omega_S)
	= \frac{i}{2\pi}\int d\omega \lim_{\eta \to 0^+}\cT^c_{ib,aj}(\omega)e^{-i\omega\eta} \qty[\frac{1}{\omega_S-\omega+\e{j}{}+\e{i}{}+i\eta}+\frac{1}{\omega_S+\omega-\e{b}{}-\e{a}{}+i\eta}] 
 \label{eq:H2preso_eigen_2_T}
\end{equation}
with $\cT^c_{ib,aj}=\cT_{ib,aj}-\mel{ib}{}{aj}$ the correlation part of $\cT$. 
Equation \eqref{eq:egval} represents a non-linear eigenvalue equation to calculate the positive excitation energies of a system, which can be rewritten as
\begin{equation}
\label{eq:H2preso_eigen}
	\sum_{jb} A_{ia,jb}(\omega_S) X_{jb,S}=\omega_S X_{ia,S}
\end{equation}
with
\begin{equation}
	A_{ia,jb}(\omega_S)=(\e{a}{}-\e{i}{})\delta_{ij}\delta_{ab} + \mel{ib}{}{aj} + \widetilde{\cT}^c_{ib,aj}(\omega_S).
\end{equation}
If one drops the dynamical part $\widetilde{\cT}^c$, one ends up with the usual time-dependent Hartree-Fock (TDHF) equations. \cite{Dreuw_2005} To calculate the correlation contribution, one can employ Eq.~\eqref{eq:T} in Eq.~\eqref{eq:H2preso_eigen_2_T}, and, after integration over the frequency, one gets Eq.~\eqref{eq:dT}.
\\
\twocolumngrid
\section*{Data availability statement}
The data that supports the findings of this study are available within the article and its supplementary material.

\bibliography{Tmatrix}

\end{document}